# More than 3-mm-long carrier diffusion and strong absorption over the full solar spectrum in copper oxide and selenium composite film

Fei Gao[a,#,*], Rongrong Gao[a,#], Hao Liu[a], Zhou Yang[a], Xiaodong Hua[a], Xin Wu[a], and Shengzhong (Frank) Liu[a,b,*]

**Abstract**: Usually, single-crystal material has a longer carrier diffusion length due to fewer defects. Here, we report a nano/amorphous copper oxide and selenium (COS) composite film prepared by radio frequency magnetron co-sputtering of CuO and Se and post annealing. The photo-generated carriers in the COS film can diffuse very far (3.147 mm). Additionally, the COS film can strongly absorb light over the entire solar spectrum region (250-2500 nm) (average absorptance is ~80%). These results indicate that the COS composite is an excellent material and can be used in solar cells and other photo-electric devices.

For materials used in solar cells and other photo-electric devices, it is important to strongly absorb light over a broad spectrum region, which produces more photo-generated carriers, and results a larger photocurrent. For electric and photo-electric materials, the other important parameter is the carrier diffusion length (CDL); a longer CDL means more carriers can be collected and better device performance can be achieved. Usually, the CDL is longer in single-crystal materials because of reduced carrier recombination resulting from fewer defects. For example, the CDLs in single-crystal p-Si[1-4], p-GaAs[5], $CH_3NH_3PbI_3$[6], and $CH_3NH_3PbBr_3$[7] are 1.95, 0.15, and 0.17, and 0.017 mm, respectively, whereas, the CDLs of poly-crystalline, nanocrystalline, and amorphous materials are very small due to a larger number of defects in the materials, which leads to significant carrier recombination. For example, the CDLs of multicrystalline Si, nanocrystalline films $CH_3NH_3PbI_{3-x}Cl_x$, and $CH_3NH_3PbI_3$ are 0.3 mm, 1.0 μm, and 0.1 μm, respectively[8,9].

Copper (Cu) has two major oxides: cupric oxide (CuO) and cuprous oxide ($Cu_2O$), both of which are native p-type semiconductors and have bandgaps of ~1.4 eV[10] and ~2.0 eV[11], respectively. CuO has a higher light absorption coefficient (~$10^6$ cm$^{-1}$). Both are promising photovoltaic materials[12-15]. The theoretical limits of the photo-electric conversion efficiencies of $Cu_2O$ and CuO solar cells estimated based on the radiative recombination mechanism are approximately 20% and 31%, respectively[16]. Of the two, the $Cu_2O$ solar cell was studied earlier[12,13], and a higher efficiency of 6.1% was achieved recently[14] for devices that benefited from a better crystal quality of the $Cu_2O$ fabricated by high temperature (~1100 °C) thermal oxidation of a Cu sheet.

Compared with $Cu_2O$ solar cells, there are fewer studies of CuO solar cells[17-19] because CuO film with high crystalline quality is difficult to prepare. Thermal evaporation cannot be used to deposit CuO film since it will decompose around its melting point (1026 °C). CuO films prepared by magnetron sputtering or via a chemical solution have a large number of defects, resulting in lower efficiencies of the corresponding solar cells (~2%). High-temperature annealing does not effectively improve the crystallinity of CuO film: besides, this is not suitable for the devices fabricated on glass substrates.

Selenium (Se) is also a p-type semiconductor material with a very high light sensitivity, its melting point is low (221 °C), and it can be prepared and processed at lower temperature. Se can be used in thermoelectric devices, photoconductors, and solar cells[20,21], etc. However, the bandgap of Se is larger (~1.9 eV), which limits its application in solar cells.

We propose a strategy to combine the advantages of Cu oxide and Se to prepare a Cu oxide and selenium (COS) composite film. The prepared COS film can strongly absorb light over the entire solar spectrum (average absorptance is ~80%). Additionally, although the prepared COS film is nano/amorphous, the photo-generated carriers have a very long diffusion length (3.147 mm), which is longer than the greatest value of both single-crystal p-Si and single-crystal $CH_3NH_3PbI_3$ perovskite.

We deposited the COS films by radio frequency magnetron co-sputtering of CuO and Se onto a 30-nm-thick compact n-type $TiO_2$ layer. The $TiO_2$ layer was prepared on transparent-conducting-fluorine-doped tin oxide (FTO)-coated glass or glass substrate by immersing the substrate in a 0.2 molL$^{-1}$ $TiCl_4$ aqueous solution for 50 min at 70 °C and then washing with distilled water and ethanol. For the sample (2 cm×7 cm) used to perform the carrier diffusion length measurement, the FTO film in a 2 cm×6 cm area was first ground away, and then the sample was dipped in 10% HF solution for 2 min to undergo a chemical polishing. The COS films were deposited on the $TiO_2$/FTO or glass substrates by radio frequency magnetron co-sputtering of CuO and Se. Ar was used as the sputtering gas, and the sputtering was performed at a pressure of 0.5 Pa, rf power of 60 W, and substrate temperature of 200 °C. After sputtering, the COS films were annealed at ~ 230 °C for 1 min in air.

We first performed X-ray diffraction (XRD) measurements to determine the material phase and crystallinity of the deposited COS films. Fig. 1a shows the XRD pattern of the as-deposited and annealed COS films deposited on the $TiO_2$/FTO glass substrates. The XRD pattern of the as-deposited COS film shows it is amorphous, excluding the FTO peaks, whereas after annealing, some new diffraction peaks from Se and $CuSe_2$ appeared. The crystallite grain sizes of the Se and $CuSe_2$ were estimated using Scherrer's formula to be 48 and 44 nm, respectively. There are no diffraction peaks from Cu oxide, reflecting it is in the amorphous state in the COS film.

X-ray photoelectron spectroscopy (XPS) measurement was further carried out to identify the chemical state of elements in the annealed COS film. As shown in Fig. 1b and (Supplementary Fig. S1b and c), the binding energies of 932.6 eV (Cu 2p3/2) and 952.6 eV (Cu 2p1/2) indicate that Cu is present in +1 oxidation state (i.e., $Cu_2O$), and the binding energies of 934.8 eV and 954.6 eV indicate that Cu is present in +2 oxidation state (i.e., CuO). The peaks around 943 and 962 eV are $Cu^{2+}$ shake up satellite peaks, which are also an evidence of CuO in the film. The content ratio of CuO to $Cu_2O$ is about 1.63 estimated by the area ratio of Cu 2p3/2 and Cu 2p1/2 peaks. In Fig. 1c (and Fig. S1d), the broad O 1s spectrum can be fitted by three peaks at 530.6, 531.2, and 532.0 eV, indicating $O^{2-}$, $O^{1-}$, and absorbed oxygen $O^0$ in the film. Se 3d spectrum can be fitted by two peaks at 53.8 and 54.7 eV, reflecting $Se_2^{2-}$ (i.e. $CuSe_2$) and $Se^0$ in the film (Fig. 1d and S1e). Therefore, the prepared COS film is a composite film of amorphous CuO, nano Se and $CuSe_2$.

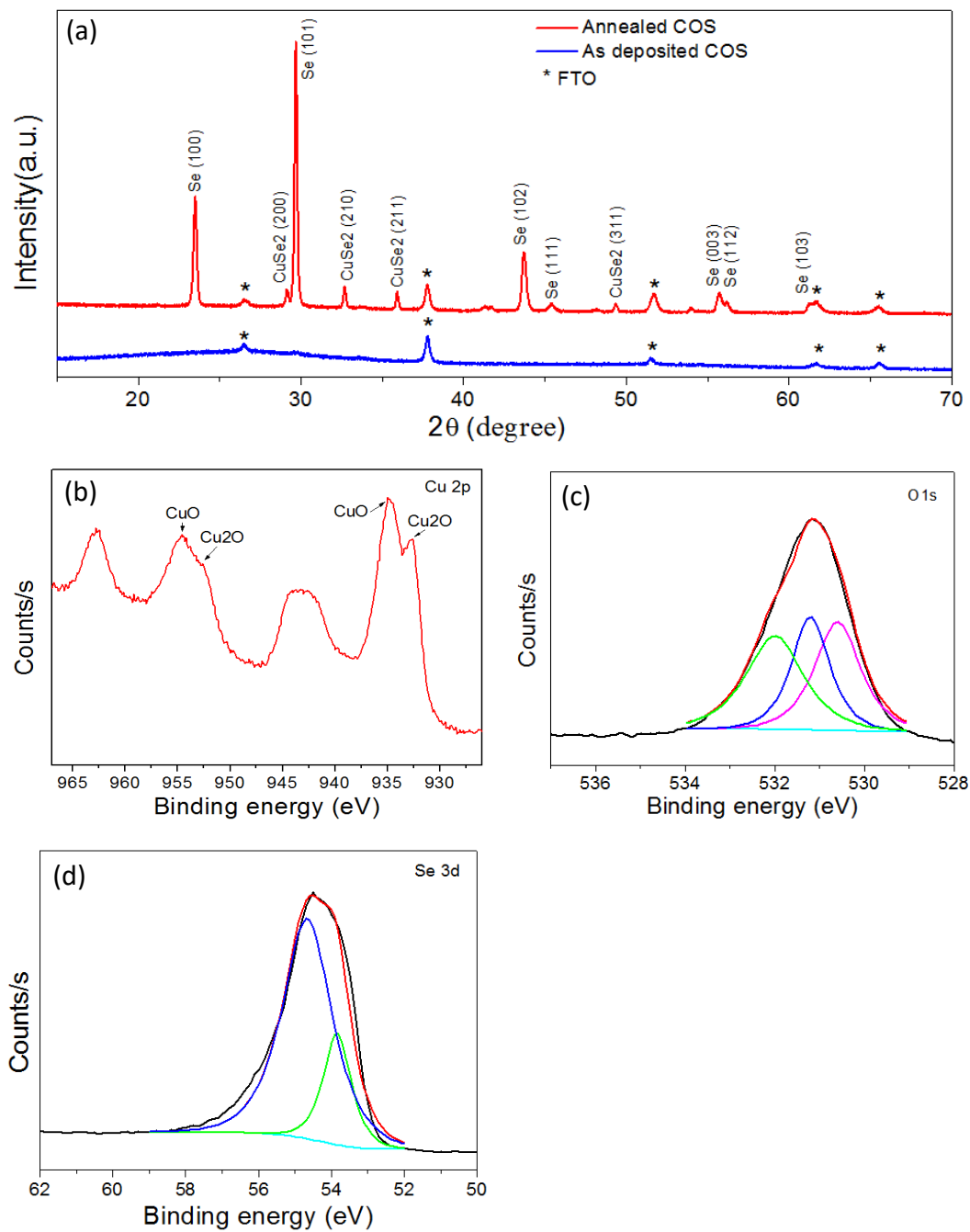

**Fig. 1**. X-ray diffraction pattern of the as-deposited and annealed COS films (a). X-ray photoelectron core level spectrum of Cu 2p (b), O 1s (c), and Se 3d (d) for the annealed COS film.

We studied the morphology of the annealed COS films by scanning electron microscopy (SEM). The top-view SEM (Fig. 2a) shows there are dense grains (~ 10-30 nm in size) on the surface of the film. The cross-sectional SEM image shows that the film is composed of large lumps, and the thickness of the film is ~ 3.9 μm (Fig. 2b). From energy-dispersive (EDX) analysis, the atomic percentages of Cu, O, and Se in the COS film are approximately 20.9%, 30.4%, and 43.3%, respectively (Fig. S2). It should be mentioned that there is ~5.4% Si in the film, which is mainly in the SiO and $Si_2O_3$ state (Fig. S1f). More fine structure of the film was revealed in the high- resolution transmission electron microscopy (HRTEM) image (Fig. 2c) and the electron diffraction pattern (Fig. 2d). These results indicate the film is a composite of crystalline and amorphous phases.

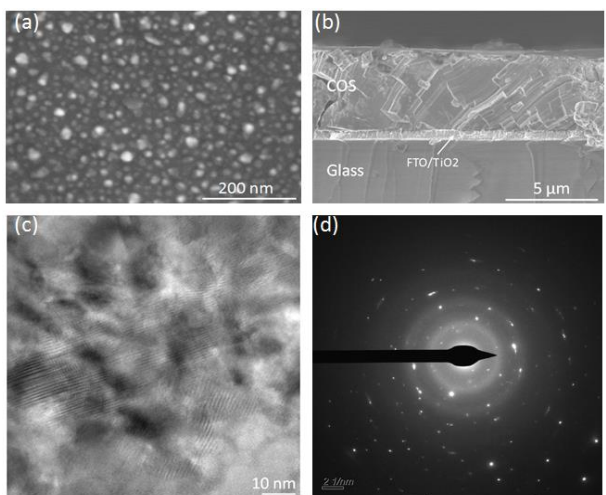

**Fig. 2.** Images of surface morphology SEM (a), cross-sectional SEM (b), cross-sectional TEM (c), and the electron diffraction pattern of the COS film (d).

During sputtering, most of the CuO and Se were directly sputtered onto the substrate, some CuO changed into $Cu_2O$ due to the loss of O in sputtering. Upon annealing, the Se melts and mixes with the CuO, some Se crystallizes and some Se reacts with Cu and produces $CuSe_2$. Thus, the annealed film is a composite of $CuO/Cu_2O$, Se, and $CuSe_2$.

The electrical properties of the COS composite film were studied by a Hall effect measurement. The film displays a p-type character with a Hall coefficient of 0.15 $cm^3$ $C^{-1}$, and its hole concentration, mobility, and resistivity are 2.05 x $10^{19}$ $cm^{-3}$, 98 $cm^2V^{-1}s^{-1}$, and 3.12 x $10^{-3}$ $\Omega$cm, respectively.

Sufficient absorption of light is very important for solar cells and other photo-electric devices. The as-deposited COS film appears reddish-black (Fig. 3a), and the annealed one is black (Fig. 3c). We studied the optical absorption properties of the COS film through ultraviolet–visible- near infrared (Uv-Vis-NIR) measurements. We measured the light transmittance $T$ and reflectance $R$ of the as-deposited and annealed COS films. The light absorptance $A$ is calculated as $A$=100%-$T$-$R$, and the $T$, $R$, and $A$ spectra are plotted in Fig. 3. The as-deposited COS film (Fig. 3b) has stronger light absorption for wavelengths below 1000 nm, corresponding to a bandgap of 1.24 eV, which is near the bandgap of CuO.

The annealed COS film (Fig. 3d) has strong light absorption (average absorptence is ~80%) in the broad spectral range from 200 to 2500 nm corresponding to the solar spectrum. (From the trend, it appears the film could absorb additional light at longer wavelengths beyond the limit of our equipment). The relative weak absorption (~52%) near 900 nm corresponds to a bandgap of 1.4 eV. Since the bandgaps of Se and $CuSe_2$ are ~1.9 and ~2.0 eV, respectively[20,22], we think the 1.4 eV bandgap of the film is mainly determined by CuO. We speculate the long wavelength (900-2500 nm) absorption is from the defect states in the bandgap of CuO[23,24], due to a large number of defects in the nano/amorphous COS film formed by sputtering.

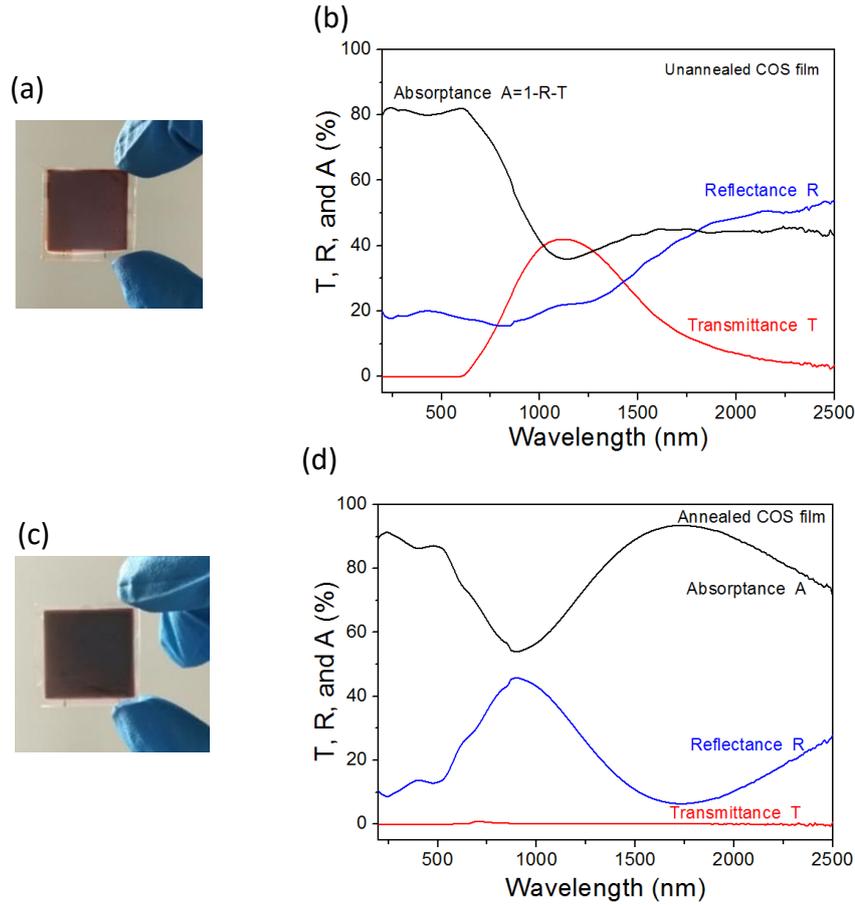

**Fig. 3**. Photographs of the un-annealed (a) and annealed (c) COS films; Transmittance *T*, Reflectance *R*, and Absorption *A* spectra of the un-annealed (b) and annealed (d) COS film.

Figure 4a shows the COS sample prepared for the measurement of the photo-generated CDL. We investigated the photo-generated CDL in our COS film by directly measuring the variation of photocurrent *I* as a function of lateral distance *L* between the illumination laser spot and the carrier collection end (carrier separation junction)[25,26], as shown in Fig. 4b and c. A He-Ne laser ($\lambda$=623.8 nm) beam passed through a chopper with a frequency of 20 Hz, a beam expander, and a rectangular aperture of dimensions 20 mm×2 mm. It illuminated the COS film with an intensity of 0.41 mW cm$^{-2}$. The sample was placed on a movable platform and was moved in 0.25-mm steps to vary the normal distance *L* between the nearest side of the carrier separation junction and the nearest edge of the laser illuminated rectangle. The photocurrent *I* was measured using a lock-in amplifier.

The separation of photo-generated electrons and holes was realized by a simple p-COS/n-TiO$_2$/FTO heterojunction. The incident light is absorbed by the COS film and excites electron and hole pairs (photo-generated carriers). The photo-generated electrons and holes diffuse and arrive at the heterojunction, where they are separated (the holes flow to the surface of COS, and the electrons flow to the FTO) and collected by the probes, producing the photo-current *I*.

The variation of *I* as a function of *L* is given in Fig. 4d. An exponential decay of *I* is observed with increasing *L* due to the recombination of the photo-generated carriers, and the equation $I = I_0 \exp(-L/L_D)$ can be used to characterize this variation[1,23,24], where $I_0$ is the photocurrent maximum when the laser spot is at the edge of the collection junction, and $L_D$ is the diffusion length of photo-generated carriers. From this equation, an extremely large $L_D = 3.147$ mm is obtained for the COS film. Although the prepared COS film is nano/amorphous, its $L_D$ is much longer than those of single-crystal p-Si (1.95 mm[1-4]), single-crystal p-GaAs (0.15 mm[5]), and single-crystal perovskite $CH_3NH_3PbI_3$ (0.175 mm[6]), and far larger than that of multi-crystalline Si (0.3 mm)[8] and nano-crystalline $CH_3NH_3PbI_{3-x}Cl_x$ thin films (1.0 μm)[9].

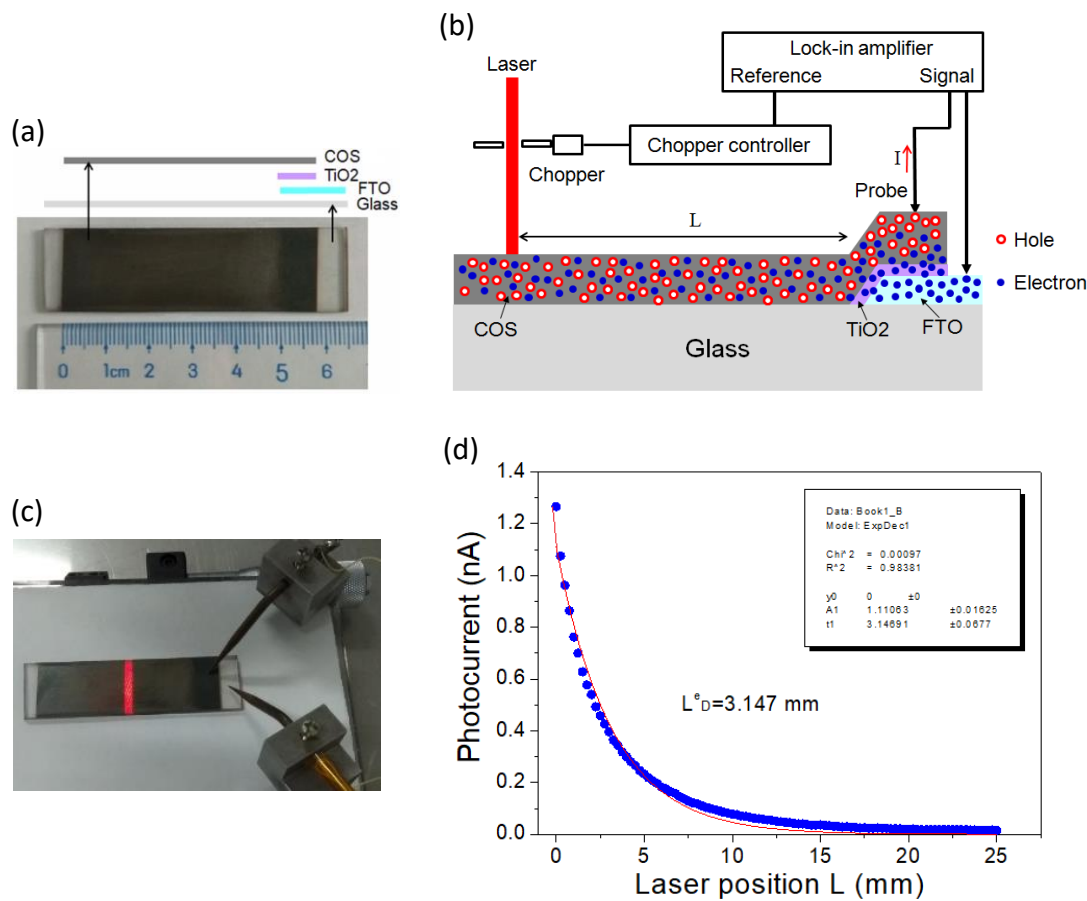

**Fig. 4**. COS sample (a), schematic diagram (b), and experimental photograph of the measurement of photo-generated CDL in the COS film (c). Photocurrent versus position *L* of the laser excitation relative to the edge of the collection junction; solid lines correspond to single exponential fits (d).

We think the mechanism that accounts for the extremely long carrier diffusion in the COS film is as follow: Since there are many defect energy states in the COS film, during the carrier diffusion process, the recombination of carriers could excite new carriers in the film. Therefore, the totality of many carriers recombination/excitation events in succession results in a long carrier diffusion length. The detailed mechanism needs to be studied further.

We have prepared a new COS film, which can strongly absorb the light over the entire solar spectrum region (average absorptance ~ 80%), and the photo-generated carriers can diffuse very far (to 3.147 mm) in the film. These excellent properties benefit from the combination of the advantages of CuO and Se, which are crucial for solar cell and other photo-electric device applications.

**Acknowledgments**
This work was supported by The National Key Research and Development Program of China (No. 2016YFA0202403; the National University Research Fund of China (GK261001009); the Chinese National 1000-talent-plan program; and the National Natural Science Foundation of China (61604091and 91733301).

**Author contributions**
F. Gao conceived and designed the experiments, performed the main experiments and data analysis, and wrote the manuscript. R.R. Gao performed the main experiments and discussed the results. H. Liu performed the deposition of some films. H. Liu and X. Wu performed the Hall measurement. Z. Yang and S.Z. Liu discussed the results and commented on the manuscript.

**Competing interests**
The authors declare no competing financial interests.